\documentclass[preprint,3p,times,twocolumn]{elsarticle}

\usepackage{amssymb}
\usepackage{amsmath}
\usepackage{amsthm}
\usepackage{graphicx}% Include figure files
\usepackage{dcolumn}% Align table columns on decimal point
\usepackage{bm}% bold math
\usepackage{color}
\usepackage{rotating}
\usepackage{setspace}
\usepackage{pseudocode}
\newcommand{\SOLVE}{\mbox{\bfseries solve }}
\newcommand{\QR}{\mbox{\bfseries QR decomposition }}
\newcommand{\INIT}{\mbox{\bfseries initial guess }}

\newcommand{\CHOOSE}{\mbox{\bfseries choose }}

\journal{Computer Physics Communications}
\newcommand{\bea}{\begin{eqnarray}}
\newcommand{\eea}{\end{eqnarray}}
\newcommand{\beaa}{\begin{eqnarray*}}
\newcommand{\eeaa}{\end{eqnarray*}}
\newcommand{\beq}{\begin{equation}}
\newcommand{\eeq}{\end{equation}}
\newcommand{\bc}{\begin{center}}
\newcommand{\ec}{\end{center}}
\newcommand{\br}{\begin{flushright}}
\newcommand{\er}{\end{flushright}}
\newcommand{\bl}{\begin{flushleft}}
\newcommand{\el}{\end{flushleft}}

\begin{document}
%\br
%\underline{Testing Block solver with SAP}\\
%\underline{Yoshifumi Nakamura}\\
%\underline{2011/Jan/26}\\
%\er

\begin{frontmatter}
\title{Modified Block BiCGSTAB for Lattice QCD}

\author{Y.~Nakamura$^{a}$, K.~-I.~Ishikawa$^{b}$, Y.~Kuramashi$^{c,d,a,f}$, T.~Sakurai$^{e,f}$, H.~Tadano$^{e,d,f}$}

\address{
 $^a$RIKEN Advanced Institute for Computational Science, Kobe, Hyogo 650-0047, Japan\\
 $^b$Graduate School of Science, Hiroshima University,   Higashi-Hiroshima, Hiroshima 739-8526, Japan\\
 $^c$Graduate School of Pure and Applied Sciences, University of Tsukuba, Tsukuba, Ibaraki 305-8571, Japan\\
 $^d$Center for Computational Sciences, University of Tsukuba, Tsukuba, Ibaraki 305-8577, Japan\\
 $^e$Department of Computer Science, University of Tsukuba,    Tsukuba, Ibaraki 305-8573, Japan\\
 $^f$JST, CREST, 5, Sanbancho, Chiyoda-ku, Tokyo 102-0075, Japan\\
}

\begin{abstract}
%We present remarkable cost reduction of multiple right-hand sides problem
%in lattice QCD by block BiCGSTAB modified by 
%the QR decomposition
% and
%domain decompostion preconditioning.
%The numerical test is done on $32^3 \times 64$ at almost physical quark mass values.
%Block size of the domains is fixed as $8\times 8 \times 8 \times 8$.
We present results for application of block BiCGSTAB algorithm 
modified by the QR decomposition and the SAP preconditioner
to the Wilson-Dirac equation with multiple right-hand sides 
in lattice QCD on a $32^3 \times 64$ lattice at almost physical quark masses.
The QR decomposition improves convergence behaviors in the block BiCGSTAB
algorithm suppressing deviation between true residual and recursive one.
The SAP preconditioner applied to the domain-decomposed lattice 
helps us minimize communication overhead.
We find remarkable cost reduction thanks to cache tuning and 
reduction of number of iterations.
\end{abstract}

\begin{keyword}
Lattice gauge theory \sep 
Lattice Dirac equation \sep 
Multiple right-hand sides \sep 
Block Krylov subspace \sep
Domain decomposition
\end{keyword}

\end{frontmatter}
%\linenumbers
\thispagestyle{plain}

\section{Introduction}
Lattice QCD simulations initiated 30 years ago 
stand finally at the point where one can obtain results of
physical observables at the physical
up, down and strange quark masses~\cite{PACSCS10}.
The next steps would be refinement of the results
reducing the systematic errors and challenge to 
computationally difficult problems, e.g. calculation of disconnected diagrams.
A main difficulty in lattice QCD simulations is that solution of Dirac equation,
which have to be repeated many times both in configuration generation 
and measurement of physical observables on given configurations, 
is computationally expensive near the physical up and down quark masses.
In the measurement of physical observables, however, 
computational cost may be reduced by block Krylov subspace methods~\cite{BCG},
since its expensive part is the multiple right-hand side problem.
(This is not the case for configuration generation.)
%It is known that the block Krylov subspace methods may be able to
%reduce the computational cost to solve such problems~\cite{BCG}.
One can expect that block Krylov subspace methods
make convergence faster with the aid of better search vectors 
generated from wider Krylov subspace enlarged by
number of right-hand side vectors
in comparison with non-blocked method.
Another possible ingredient to improve performance
is an efficient use of memory bandwidth 
in implementation of block matrix-vector multiplication.

Since the Dirac matrix in lattice QCD is non-Hermitian,
we might expect the block BiCGSTAB algorithm~\cite{BBCGSTAB} 
is applicable in a straightforward way.
One problem in block Krylov subspace methods, however, is that 
the true residual stops decreasing at some point, while
the recursive one continues to decrease.
Recently, three of the authors have proposed a new algorithm named 
block BiCGGR, which showed significant improvement 
for this problem~\cite{BBiCGGR1,BBiCGGR2,BBiCGGR3}.
%In these papers, two problems which are the deviation between the true and the recursive 
%residuals and the covergence at smaller quark masses are improved.

In this paper we improve block BiCGSTAB algorithm with two modifications. 
First one is the QR decomposition, which is known to improve 
the numerical accuracy in block CG~\cite{BCG,GSBCG} and
also useful for block BiCGSTAB algorithm~\cite{GSBiCGSTAB_theo}.
Second one is Schwarz alternating procedure (SAP) 
preconditioner proposed by L\"uscher~\cite{SAP}, which
is applied to the domain-decomposed lattice.
We can minimize communication overhead with the SAP preconditioner.

This paper is organized as follows. In Sec.~\ref{sec:algorithm} we explain
the algorithmic details of the modified block BiCGSTAB with SAP 
preconditioner. We present the results of the
numerical test in Sec.~\ref{sec:test}. Conclusions and discussions are
summarized in Sec.~\ref{sec:conclusion}. 

%%%%%%%%%%%%%%%%%%%%%%%%%%%%%%%%%%%%%%%%%%%%%%%%%%%%%%
\section{Algorithm}\label{sec:algorithm}
\subsection{Modified Block BiCGSTAB}
We consider to solve the linear systems with the multiple right-hand sides expressed as
\begin{equation}
AX=B\,,
\end{equation}
where $A$ is an $N\times N$ complex sparse non-Hermitian matrix.
$X$ and $B$ are $N\times L$ complex rectangular matrices given by
\bea
X&=&\left(\bm{x}^{(1)},\dots,\bm{x}^{(i)},\dots,\bm{x}^{(L)} \right),\\
B&=&\left(\bm{b}^{(1)},\dots,\bm{b}^{(i)},\dots,\bm{b}^{(L)} \right).
\eea
In the case of the Wilson-Dirac equation the matrix dimension is
given by
$N=L_x\times L_y\times L_z\times L_t \times 3 \times 4$
with $L_x\times L_y\times L_z\times L_t$ the volume of 
a hypercubic four-dimensional lattice.
$L$ is the number of the right-hand-side vectors which is called
source vectors in lattice QCD. $L$ is 12 in the simplest case and 
$O(10)-O(100)$ (perhaps more in some case) for the stochastic method.

The matrix-vector multiplication for the Wilson-Dirac equation is written as
\begin{equation}
 A \phi = \sum_{x=1}^{L_x\times L_y\times L_z\times L_t} (\phi_x-\kappa \eta_x)\,,
\end{equation}
\begin{equation}
 \eta_x = \sum_{\mu=1}^4 \Big[(1-\gamma_\mu) U_{x,\hat{\mu}} \phi_{x+\hat{\mu}} +
                        (1+\gamma_\mu) U^\dagger_{x-\mu,\hat{\mu}} \phi_{x-\hat{\mu}} \Big]\,,
\label{eq:hoppingterm}
\end{equation}
where $\phi_x$ and $\eta_x$ contain 12 complex numbers at site $x$, $\gamma_\mu$ are the gamma matrices,
$U_{x,\hat{\mu}}$ are link variables of SU(3) matrix and $\kappa$ is hopping parameter.
Computation of $\eta_x$ requires 
1320\footnote{1296 flops if $\gamma_\mu$ is non-relativistic representation.}
floating point number operations and 360 words per site.
%\footnote{24 words for store and 336 words for load.
%144 of 336 words can be reconstructed from less words since they are components of eight $SU(3)$ matrices
%which are connecting to a site.} 
%words per site. 
This means the value of Flops/Byte is about 0.9 (0.45) 
in the single (double) precision.
It should be difficult to obtain high performance 
on recent computer architecture.

\begin{figure}[!t]
%\begin{center}
\includegraphics[width=60mm,angle=-90]{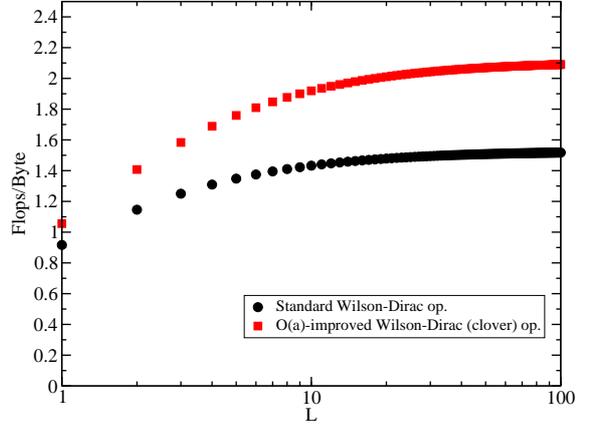}
%\end{center}
\vspace{-.5cm}
\caption{Flops/Byte as a function of $L$ for standard (black circle) and 
$O(a)$-improved (red square) Wilson-Dirac operators in the single precision.}
\label{fig:flopsbyte}
\end{figure}

In block Krylov subspace methods, Eq.~(\ref{eq:hoppingterm}) can be expressed as
\begin{equation}
\begin{split}
%\left(\bm{\eta_x}^{(1)},\dots,\bm{\eta_x}^{(L)} \right)&= \sum_{\mu=1}^4 \Big[
%(1-\gamma_\mu) U_{x,\hat{\mu}} \left(\bm{\phi_{x+\hat{\mu}}}^{(1)},\dots,\bm{\phi_{x+\hat{\mu}}}^{(L)} \right)\\
%+&(1+\gamma_\mu) U^\dagger_{x-\mu,\hat{\mu}} 
%\left(\bm{\phi_{x-\hat{\mu}}}^{(1)},\dots,\bm{\phi_{x-\hat{\mu}}}^{(L)} \right) \Big]\,.
\bm{\eta_x}^{(i)}= \sum_{\mu=1}^4 \Big[
(1-\gamma_\mu) U_{x,\hat{\mu}} \bm{\phi_{x+\hat{\mu}}}^{(i)}
+(1+\gamma_\mu) U^\dagger_{x-\mu,\hat{\mu}}\bm{\phi_{x-\hat{\mu}}}^{(i)} \Big]\,,
\label{eq:bloclhoppingterm}
\end{split}
\end{equation}
with $i=1,\dots,L$.
An important point is that same 8 link variables around site $x$ 
are used in common for $\phi^{(i)}$ with $i=1,\dots,L$ and their size is
just 576 (1152) bytes in the single (double) precision, which 
are small enough to be retained in low level cache, 
for example L1 cache (if there is).
This allows us more efficient usage of cached data than 
repeating $L$ independent matrix-vector multiplications.
Figure~\ref{fig:flopsbyte} illustrates how 
Flops/Byte increases as $L$ is enlarged. 
For an effective use of cache, we arrange loop for the index of $i$ 
($i=1,\dots,L$) in the most inner level 
with $i$ running first in memory allocation for vectors.

Pseudocode for modified block BiCGSTAB algorithm is described in 
Algorithm~\ref{alg:bbs}.
Note that preconditioner $M$ at lines 4.2 and 4.6 in Algorithm~\ref{alg:bbs}
must be implemented by a stationary iterative method in this algorithm
%since the same matrix should be multiplied for all the vectors, though
since the common preconditioning should be applied to all right-hand sides, though
nonstational iterative methods are often used in lattice QCD~\cite{mix2}.
The orthogonalization of $P$ improves numerical accuracy
since each span works effectively to search approximated solutions.
We employ modified Gram-Schmidt method for the QR decomposition.
%Linear independence of each span also works effectively against ill-conditioned problem.
Even when non-block BiCGSTAB fails to converge, modified block BiCGSTAB may converge 
by adding dummy right-hand side vectors if they can play 
a supplementary role~\cite{GSBiCGSTAB_theo}.
We also present a memory saving version in Algorithm~\ref{alg:bbs_memsave}.

%\noindent
%\underline{Modified (orthogonalized) block BiCGSTAB}\\
%\noindent
%Compute: $R_0=B-AX_0$\\
%Set: $P_0=R_0$\\
%Choose: $\tilde{R}$\\
%For k=0,1,..., until $\max|r|< \epsilon |b|$ do: \\
%\hspace{5mm}Orthogonalize: $P_k$ \\
%\hspace{5mm}Precondition: $U_k= M P_k$ \\
%\hspace{5mm}$V_k = A U_k$ \\
%\hspace{5mm}Solve: $(\tilde{R}^H V_k) \alpha_k = \tilde{R}^H R_k$ \\
%\hspace{5mm}$T_k=R_k-V_k \alpha_k$\\
%\hspace{5mm}Precondition: $S_k=MT_k$\\
%\hspace{5mm}$Z_k=AS_k$\\
%\hspace{5mm}$\zeta_k = \mbox{Tr}(Z^H_k T_k)/\mbox{Tr}(Z^H_k Z_k)$ \\
%\hspace{5mm}$X_{k+1}= X_k+U_k \alpha_k +\zeta_k S_k$\\
%\hspace{5mm}$R_{k+1} = T_k - \zeta_k Z_k$, \\
%\hspace{5mm}Solve: $(\tilde{R}^H V_k)\beta_k = - \tilde{R}^H Z_k$ \\
%\hspace{5mm}$P_{k+1}= R_{k+1} +(P_k- \zeta_k V_k)\beta_k$\\
%end For:\\

\begin{figure}
{\small 
\begin{spacing}{0.75}
\begin{pseudocode}{Modified Block BiCGSTAB}{A,M,B,\epsilon}
1~~~~ \INIT X \in \mathbb{C}^{N \times L}\\
2~~~~ R = B - AX\\
3~~~~ P = R\\
4~~~~ \CHOOSE \tilde{R} \in \mathbb{C}^{N \times L}\\
\WHILE \max_i(| \bm{r}^{(i)}|/ | \bm{b}^{(i)}|) \le \epsilon \DO
\BEGIN
 4.1~~~~  \QR P=Q \gamma,~P \GETS Q \\
 4.2~~~~  U = M P \\
 4.3~~~~  V = A U \\
 4.4~~~~  \SOLVE (\tilde{R}^H V) \alpha = \tilde{R}^H R~~\FOR \alpha \\
 4.5~~~~  T = R - V \alpha \\
 4.6~~~~  S = M T \\
 4.7~~~~  Z = A S \\
 4.8~~~~  \zeta = \mbox{Tr}(Z^H_k T_k)/\mbox{Tr}(Z^H_k Z_k) \\
 4.9~~~~  X \GETS X + U \alpha  + \zeta S \\
 4.10~~ R = T - \zeta Z \\
 4.11~~ \SOLVE (\tilde{R}^H V)\beta = - \tilde{R}^H Z~~\FOR \beta \\
 4.12~~ P \GETS R + ( P - \zeta V)\beta \\
\END \\
5~~~~ \RETURN{X}
\label{alg:bbs}
\end{pseudocode}
\end{spacing}
}
\end{figure}
%\footnote{To save memory one can use memory space of $R$ for $T$.}

\begin{figure}
{\small 
\begin{spacing}{0.75}
\begin{pseudocode}{Memory Saving Version}{A,M,B,\epsilon}
1~~~~  \INIT X \in \mathbb{C}^{N \times L}\\
2~~~~  R = B - AX\\
3~~~~  P = R\\
4~~~~  \CHOOSE \tilde{R} \in \mathbb{C}^{N \times L}\\
\WHILE \max_i(| \bm{r}^{(i)}|/ | \bm{b}^{(i)}|) \le \epsilon \DO
\BEGIN
 4.1~~~~  \QR P=Q \gamma,~P \GETS Q \\
 4.2~~~~  U = M P \\
 4.3~~~~  V = A U \\
 4.4~~~~  \SOLVE (\tilde{R}^H V) \alpha = \tilde{R}^H R~~\FOR \alpha \\
 4.5~~~~  R \GETS R - V \alpha \\
 4.6~~~~  X \GETS X + U \alpha \\
 4.7~~~~  S = M R \\
 4.8~~~~  Z = A S \\
 4.9~~~~  \zeta = \mbox{Tr}(Z^H_k R_k)/\mbox{Tr}(Z^H_k Z_k) \\
 4.10~~ X \GETS X  + \zeta S \\
 4.11~~ R \GETS R - \zeta Z \\
 4.12~~ \SOLVE (\tilde{R}^H V)\beta = - \tilde{R}^H Z~~\FOR \beta \\
 4.13~~ P \GETS R + ( P - \zeta V)\beta \\
\END \\
5~~~~ \RETURN{X}
\label{alg:bbs_memsave}
\end{pseudocode}
\end{spacing}
}
\end{figure}

%\begin{quote}
%   $X_0 \in \mathbb{C}^{n \times L}$ is an initial guess, \\
%   {\bf Compute} $R_0 = B - AX_0$, \\
%   {\bf Set} $P_0 = R_0$, \\
%   {\bf Choose} $\tilde{R}_0 \in \mathbb{C}^{n \times L}$, \\
%   {\bf For} $k=0, 1, \dots, $ {\bf until}
%   $\max_i(| \bm{r}_k^{(i)}|/ | \bm{b}^{(i)}|)\le \epsilon$ {\bf do:} \\
%   \begin{tabular}{lcl}
%    \multicolumn{3}{l}{{\bf Perfrom} the QR decomposition $P_k=Q \gamma$ then $P_k=Q$,} \\
%    \multicolumn{3}{l}{{\bf Preconditioning} $U_k= M P_k$,} \\
%    $V_k$ &=& $A U_k$, \\
%    \multicolumn{3}{l}{{\bf Solve} $(\tilde{R}^H V_k) \alpha_k = \tilde{R}^H R_k$ for $\alpha_k$,} \\
%    $T_k$ &=& $R_k-V_k \alpha_k$, \\
%    \multicolumn{3}{l}{{\bf Preconditioning} $S_k= M T_k$,} \\
%    $Z_k$ &=& $AS_k$,\\
%    $\zeta_k$ &=& $\mbox{Tr}(Z^H_k T_k)/\mbox{Tr}(Z^H_k Z_k)$, \\
%    $X_{k+1}$ &=& $X_k+U_k \alpha_k +\zeta_k S_k$,\\
%    $R_{k+1}$ &=& $T_k - \zeta_k Z_k$, \\
%    \multicolumn{3}{l}{{\bf Solve} $(\tilde{R}^H V_k)\beta_k = - \tilde{R}^H Z_k$ for $\beta_k$,} \\
%    $P_{k+1}$ &=& $R_{k+1} +(P_k- \zeta_k V_k)\beta_k$\\
%   \end{tabular}\\
%   {\small \bf End}
%\end{quote}

%%%%%%%%%%%%%%%%%%%%%%%%%%%%%%%%%%%%%%%%%%%%%%%%%%%%%%
\subsection{Preconditioning}

In this work we employ the $O(a)$-improved
\footnote{The leading cut-off error in terms of the lattice spacing $a$ is removed.} Wilson fermions.
After Jacobi preconditioning (division of $I$ + clover term),
the matrix $A$ is decomposed as 
the following $2 \times 2$ blocked matrix form,
\begin{equation}
A = 
 \left( \begin{array}{cc}
 A_{EE} & A_{EO} \\
 A_{OE} & A_{OO} \\
 \end{array} \right)\,,
\end{equation}
where the subscript $E$ and $O$ denote the even and odd domains, respectively. 
The SAP preconditioner $M_{SAP}$ is computed as
\begin{equation}
\begin{split}
M_{SAP}&=K \sum_{j=0}^{N_{SAP}} (1-AK)^j \,, \\
 K & = 
  \left( \begin{array}{cc}
   B_{EE}               & 0      \\
  -B_{OO} A_{OE} B_{EE} & B_{OO} \\
  \end{array} \right)\,, \\
\end{split}
\end{equation}
where $B_{EE}$ ($B_{OO}$) is an approximation for $A^{-1}_{EE}$ ($A^{-1}_{OO}$)
obtained by SSOR method~\cite{SSOR}
\begin{equation}
 B_{EE} = (1-\omega U_{EE})^{-1} \Big[ \sum_{j=0}^{N_{SSOR}} (1-A_{EE}^{SSOR})^j \Big] 
 (1-\omega L_{EE})^{-1}\,,
\end{equation}
with
\begin{equation}
\begin{split}
 A_{EE}^{SSOR} &= {1\over \omega} \Big[ (1-\omega L_{EE})^{-1} + (1-\omega U_{EE})^{-1} \\
               &\quad + (\omega -2 )(1-\omega L_{EE})^{-1}(1-\omega U_{EE})^{-1} \Big]\,.
\end{split}
\end{equation}
$L_{EE}$ is the forward hopping term and $U_{EE}$ is the backward one.
We perform SAP preconditioning in the single precision 
for effective use of memory bandwidth and network bandwidth
between nodes.

It is known that ``sloppy'' precision can be used 
in right preconditioning, but not in left one.
Suppose calculation of $S=MT$ at line 4.6 
in Algorithm~\ref{alg:bbs}
is performed with ``sloppy'' precision in $k$-th iteration.
Numerical errors for $S_k$, $Z_k$, $\zeta_k$ and $X_{k+1}$ may be expressed as
\begin{eqnarray}
S_k     &\rightarrow& S'_k =S_k  + \delta S_k \,,\\
%\end{equation}
%\begin{equation}
Z_k     &\rightarrow& Z'_k =AS'_k  \,,\\
%\end{equation}
%\begin{equation}
\zeta_k &\rightarrow& \zeta'_k =\zeta_k +\delta\zeta_k \,,\\
%\end{equation}
%\begin{equation}
X_{k+1} &\rightarrow& X'_{k+1} =X_k +U_k  \alpha_k +\zeta'_k S'_k \,.
\end{eqnarray}
These yield
\begin{equation}
\begin{split}
R'_{k+1}&= R_k-V_k \alpha_k - \zeta'_k Z'_k\\
        &= R_k-AU_k \alpha_k - \zeta'_k AS'_k \\
        &= B - AX_k -A (U_k \alpha_k + \zeta'_k S'_k)\\
        &= B - AX'_{k+1}\,, \\
\end{split}
\end{equation}
which satisfies the exact relation between approximate solutions 
and residuals in $(k+1)$-th iteration. 
For the case that both $U=MP$ at line 4.2 and $S=MT$ at line 4.6 
are computed with ``sloppy'' precision 
one can also reproduce the above relation with the following formulae:
\begin{eqnarray}
U_k     &\rightarrow& U'_k =U_k  + \delta U_k \,,\\
V_k     &\rightarrow& V'_k =AU'_k  \,,\\
\alpha_k&\rightarrow& \alpha'_k = \alpha_k + \delta \alpha\,,\\
T_k     &\rightarrow& T'_k =R_k -V'_k \alpha'_k\,,\\
S_k     &\rightarrow& S''_k= S_k +\delta S\,,\\
Z_k     &\rightarrow& Z''_k =AS''_k  \,,\\
\zeta_k &\rightarrow& \zeta''_k=\zeta_k +\delta \zeta \,,\\
X_{k+1} &\rightarrow& X''_{k+1} =X_k +U'_k  \alpha'_k +\zeta''_k S''_k \,.
\end{eqnarray}

%%%%%%%%%%%%%%%%%%%%%%%%%%%%%%%%%%%%%%%%%%%%%%%%%%%%%%
\section{Numerical test}\label{sec:test}
\subsection{Choice of parameters}
We test modified block BiCGSTAB employing a so-called ``local source'', 
$B = [e_1,...,e_L]$, with $L=12$ for color-spin components.
We use statistically independent 10 configurations generated
at almost the physical point, $\kappa_{ud}=0.137785$ and $\kappa_{s}=0.136600$,
in 2+1 flavor lattice QCD with the nonperturbatively $O(a)$-improved Wilson quark action and 
the Iwasaki gauge action~\cite{iwasaki} at $\beta=1.9$ on a $32^3 \times 64$ lattice~\cite{PACSCS10}.
We choose the hopping parameter $\kappa=0.137785$ for the Wilson-Dirac equation and
$N_{SAP}=5$ with $8 \times 8 \times 8 \times 8$ domain size
for the SAP preconditioning following Ref.~\cite{PACSCS10}.
Parameters for SSOR method are also fixed with $N_{SSOR}=1$ and $\omega=1.26$.
The stopping criterion is set to be  
$\max_i(| \bm{r}^{(i)}|/ | \bm{b}^{(i)}|) \le \epsilon$ with $\epsilon=10^{-14}$.

\subsection{Test environment}
Numerical test is performed on 16 nodes of a large-scale cluster system called T2K-Tsukuba.
The machine consists of 648 compute nodes providing 95.4Tflops of computing
capability. Each node consists of quad-socket, 2.3GHz Quad-Core AMD
Opteron Model 8356 processors whose on-chip cache sizes are 64KBytes/core,
512KBytes/core, 2MB/chip for L1, L2, L3, respectively. Each processor has
a direct connect memory interface to an 8GBytes DDR2-667 memory and
three hypertransport links to connect other processors. All the nodes in
the system are connected through a full-bisectional fat-tree network consisting
of four interconnection links of 8GBytes/sec aggregate bandwidth with Infiniband.
For numerical test we modify the lattice QCD simulation program
LDDHMC/ver1.04b12.31 developed by PACS-CS Collaboration~\cite{LDDHMC}.

%\subsection{miscellaneous}
%Intel(R) Fortran Intel(R) 64 Compiler Professional for applications running on Intel(R) 64, Version 11.1\\
%No SSE, No prefetch\\
%FFLAGS  = -O2 -xO -align all -inline-max-total-size=10000 -stand f95 -u -i4 -r8 -ip -free -fpe0\\
%mpirun\_rsh -np 16 \\
%OMP\_NUM\_THREADS=1\\

%%%%%%%%%%%%%%%%%%%%%%%%%%%%%%%%%%%%%%%%%%%%%%%%%%%%%%
\subsection{Results}

%Fig.~\ref{nocache} is typical case for residual v.s. iteration with parameter ($N_{mr}$,$N_{cy}$).
%For smaller ($N_{mr}$,$N_{cy}$) solver does not converge due to blowing-up.
%In table~\ref{nocache}, we show ratio of failure, total time to calculate inversion of $D$ 
%for all 12 colour-spin components at one local source, gain factor of time compared with $L=1$
%at same $N_{cy}$, number of accumurated iteration, number of matrix-vector multiplication (NMVM)
%and its gain factor for $L=1,2,3,4,6,12$ and $N_{cy}=5,10,15,20$ on small lattice.

%Convergece at large $L$ is improved by increasing $N_{cy}$, althought the gain is not significant.
%Time and NMVM gain is almost same due to no cache tuning.

\begin{figure}[!t]
\begin{center}
\includegraphics[width=60mm,angle=-90]{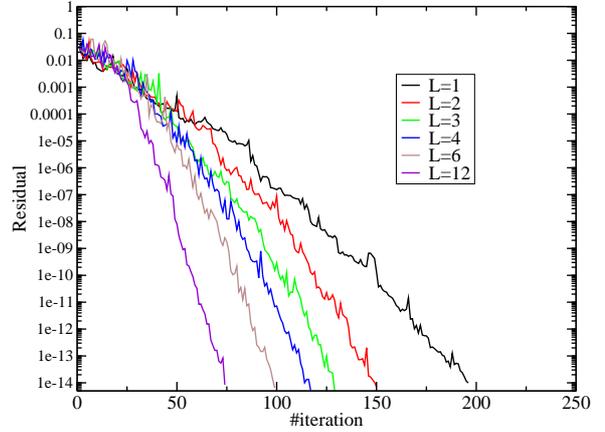}
\end{center}
\vspace{-.5cm}
\caption{Representative case for residual norm as a function of 
number of iteration with $L=1,2,3,4,6,12$.}
\label{fig:resiter}
\end{figure}

Figure \ref{fig:resiter} shows a representative case for residual norm as a function of number of iterations for modified block BiCGSTAB.
We observe one of important features of block Krylov subspace methods
that the number of iterations required for convergence decreases 
as the block size $L$ is increased.   
%The residual norms reach $10^{-14}$ for all $L$.

\begin{table}[!h]
\begin{center}
{\footnotesize
\begin{tabular}{c|c|c|c|c} \hline
$L\times 12/L$ &   time[s] & T(gain)&       NMVM &NM(gain)\\ \hline
% 1\times 12 & 3827(755) & 1      & 17146(3326)& 1      \\
% 2\times 6 & 2066(224) & 1.8(2) & 12942(1379)& 1.3(2) \\
% 3\times 4 & 1619(129) & 2.4(4) & 10652(832) & 1.6(3) \\
% 4\times 3 &  1145(99) & 3.3(5) &  9343(835) & 1.8(3) \\
% 6\times 2 &  1040(87) & 3.7(6) &  7888(663) & 2.2(4) \\
%12\times 1 &   705(70) &   5(2) &  6106(633) & 2.8(6) \\ \hline
 $1\times 12$ & 3827(755) & 1   & 17146(3326)& 1      \\
 $2\times 6$  & 2066(224) & 1.9 & 12942(1379)& 1.3 \\
 $3\times 4$  & 1619(129) & 2.4 & 10652(832) & 1.6 \\
 $4\times 3$  &  1145(99) & 3.3 &  9343(835) & 1.8 \\
 $6\times 2$  &  1040(87) & 3.7 &  7888(663) & 2.2 \\
 $12\times 1$ &   705(70) & 5.4 &  6106(633) & 2.8 \\ \hline
\end{tabular}
}
\end{center}
\vspace{-5mm}
\caption{$L$ dependence for time, gain factor of time, 
number of matrix-vector multiplication and its gain factor. Central values
are given for gain factors.}
\label{tab:gsbcstab}
\end{table}

The results are summarized in Table~\ref{tab:gsbcstab}. Second column is 
total time to solve the Wilson-Dirac equation for all 12 colour-spin 
components at one local source.
In case of $L=6$, for example, 12 right-hand side vectors are divided 
into two blocks: $B_1 = [e_1,...,e_6]$ and $B_2 = [e_7,...,e_{12}]$.
Third column is gain factor of time compared with $L=1$ case. Fourth and 
fifth columns are number of matrix-vector multiplication (NMVM) and its gain factor, respectively.
Modified block BiCGSTAB with $L=12$ is about 5 times faster than $L=1$ case. 
The iteration number is reduced by a  
factor of three. Additional speed-up by a factor of two is obtained 
by cache tuning.
This is roughly consistent with the enhancement of Flops/Byte from 1.05 
at $L=1$ to 1.95 at $L=12$
plotted in Fig.~\ref{fig:flopsbyte}.

%%%%%%%%%%%%%%%%%%%%%%%%%%%%%%%%%%%%%%%%%%%%%%%%%%%%%%
\section{Conclusions}\label{sec:conclusion}
In this paper, we have carried out numerical test for block BiCGSTAB with 
two modifications of the QR decomposition and the SAP preconditioner 
in lattice QCD at almost physical quark masses.
The QR decomposition successfully removes the problem in block BiCGSTAB
that is the deviation between the true and the recursive residuals.
We find remarkable cost reduction at large $L$ due to smaller 
number of iterations and efficient cache usage.
Further gain could be expected in calculations of disconnected diagram and 
reweighting factor, where larger value of $L$ is required.
One concern is that numerical cost for modified Gram-Schmidt method 
increases as $O(L^2)$.

%Easy wolkaround is performing modified Gram-Schmidt method every second iteration, or so.
%Block BiCGSTAB + the QR decomposition could solve very ill-conditioned case~\cite{GSBiCGSTAB_theo}.

\section*{Acknowledgments}
Numerical calculations for the present work have been carried out
on the T2K-Tsukuba computer at the University of Tsukuba.
%under the ``Interdisciplinary Computational Science Program'' of
%Center for Computational Sciences, University of Tsukuba.
This work is supported in part by Grants-in-Aid for Scientific Research
from the Ministry of Education, Culture, Sports, Science and Technology
(Nos.
%???,        %Sakurai
20800009,   %Tadano 
20105002,   %Kuramashi_shingakujyutsu
22244018    %Kuramashi_kiban_A
).
%\end{acknowledgments}

\end{document}